\documentclass[prd,twocolumn,preprintnumbers]{revtex4}
\usepackage{amsfonts}
\usepackage{amsmath}
\usepackage{graphicx}
\usepackage{dcolumn}
\usepackage{bm}
\usepackage{amssymb}

\begin{document}
\title{Electric charge in the field of a magnetic event in three-dimensional spacetime}
\author{Claudio Bunster and Cristi{\'a}n Mart\'{\i}nez  \vspace{1mm}}

\affiliation{Centro de Estudios Cient\'{\i}ficos (CECs), Av.  Arturo Prat
 514, Valdivia, Chile \\ and Universidad Andr\'{e}s Bello, Av.
Rep\'{u}blica 440, Santiago, Chile}  
 

 \vspace{15mm}

\begin{abstract}

 We analyze the motion of an electric charge in the field of a magnetically charged event in three-dimensional spacetime. We start by exhibiting a first integral of the equations of motion in terms of the three conserved components of the spacetime angular momentum, and then proceed numerically. After crossing the light cone of the event, an electric charge initially at rest starts rotating and slowing down. There are two lengths appearing in the problem: (i) the characteristic length $\frac{q g}{2 \pi m}$, where $q$ and $m$ are the electric charge and mass of the particle, and $g$ is the magnetic charge of the event; and (ii) the spacetime impact parameter $r_0$. For $r_0 \gg \frac{q g}{2 \pi m}$, after a time of order $r_0$, the particle makes sharply a quarter of a turn and comes to rest at the same spatial position at which the event happened in the past. This jump is the main signature of the presence of the magnetic event as felt by an electric charge. A derivation of the expression for the angular momentum that uses Noether's theorem in the magnetic representation is given in the Appendix.

\end{abstract}



\maketitle

\section{Introduction} \label{Introduction}

The present article deals with the motion of an ordinary electrically charged particle in the background of a magnetic point source in three-dimensional Minkowski spacetime. Such a source is the simplest case of what has been called recently a ``magnetic event," a novel  notion that emerges from the principle of electric-magnetic duality, and has been 
developed in \cite{BBH,BGH}. As it was discussed in \cite{BBH}, this case is of interest,  not only because of its simplicity,  but also because it is realized in the laboratory when an electric charge crosses a Josephson junction separating two superconducting bulk pieces.

To our knowledge this problem has not been discussed before, perhaps because a cogent  physical motivation did not yet exist.

On the other hand, the corresponding problem of determining the spatial trajectories of an electrically charged particle in the field of a magnetic monopole in three-dimensional Euclidean space (i.e., in \textit{four} dimensional Minkowski spacetime) was first considered by Poincar{\'e} over 100 years ago \cite{Poincare}, and there are many more recent treatments (see, for example \cite{Milton}). In the Euclidean case the spatial trajectories are spirals on a cone whose aperture is determined by the angular momentum. The cone becomes a plane when the magnetic pole is absent.

 At the beginning of the present study, the authors felt that, perhaps, there would be a way of directly translating in a simple manner, by some sort of ``Wick rotation,"  the  well-analyzed Euclidean results to the case of Lorentzian signature.  However efforts in this direction proved unsuccessful. Therefore it was decided to attack the problem directly, \textit{ab initio} in Minkowski space, and, as it will be seen below, the results that emerged appear to bear no evident relationship with those of the Euclidean case. 
  
With hindsight, from a physical point of view the lack of such a relationship should not be that surprising but rather is to be expected. Indeed, as emphasized in \cite{BBH,BGH}  the magnetic pole in three-dimensional Euclidean space corresponds to an ``instanton" \cite{thooft}, or ``pseudo particle" \cite{Polyakov} and describes quantum mechanical tunneling within the three-dimensional spacetime, while, in contradistinction,  magnetic events occur  classically and correspond  to the imprint of a flux-carrying particle impinging from extra dimension.

The plan of the paper is as follows. Section \ref{sec2} recalls the field of a magnetic event. Next,  Sec. \ref{sec3} discusses a first integral of the equation of motion, in which the integration constants are the three conserved components of the angular momentum. Then Sec. \ref{sec4} is devoted to further complete integration of the equations of motion, for which we resort to numerical methods. It is observed that one must regularize the otherwise divergent field of the event on the light cone. The main conclusions obtained are independent of regularization and they are stated in Sec. \ref{conclusion}:  As the electric charge crosses the light cone of the event it starts rotating  and slowing down. In the frame in which the particle is initially at rest,  after a time of the order of the spacetime impact parameter, the particle makes sharply a quarter of a turn and comes to rest at the same spatial position at which the event happened in the past. This jump is the main signature of the presence of the magnetic event as felt by an electric charge. A derivation of the expression for the angular momentum that uses Noether's theorem in the magnetic representation is given in the Appendix.

\section{Field of a magnetic event} \label{sec2}

The electromagnetic field strength $F_{\mu \nu}$ is given by
\begin{equation} \label{strength}
F_{\mu \nu}= \epsilon_{\mu\nu\rho} {}^{*}F^{\rho},
\end{equation}
where the dual ${}^{*}F^{\rho}$ is related to the magnetic potential $\varphi$ by
\begin{equation} \label{strengthphi}
{}^{*}F_{\mu}= \partial_{\mu} \varphi.
\end{equation}
If the event is at the origin $x=0$, the equation of motion for $\varphi$ reads, 
\begin{equation} \label{field}
\square\varphi=g\delta^{(3)}(x)\ . 
\end{equation}
The general solution of\ (\ref{field}) is the sum  $\varphi
=\varphi_{0}+gG$, where $\varphi_{0}$ is the general solution of the
homogeneous equation $(g=0)$ and $G$ is a Green function of the wave operator.
At the classical level a natural choice is to take for $G$ the retarded Green
function which vanishes outside the future  light cone of the event,
\begin{equation}
G_{R}(x)=\frac{1}{2\pi}\frac{1}{\sqrt{-x_{\mu}x^{\mu}}}\theta
(-x_{\mu}x^{\mu})\theta(x^{0}).  \label{Green}
\end{equation}
 If one takes $\varphi_0 = 0$ 
the situation as seen
from within the  $D=3$ spacetime  is the following:
For $x^{0}<{0}$ there is no field.
\ At $x^0 = 0$, suddenly a flash of   light emerges and propagates to
the future. As argued in \cite{BBH} this situation will be the most probable classically, because it corresponds
to  a flux-carrying particle impinging from
the extra dimension without any precise control  (``fine tuning") 
of its initial conditions.   The time-reversed process, 
where one would replace the retarded Green function
$G_{R}$ by the advanced one $G_{A},$ corresponds  to  a precisely 
prepared (``fine tuned") pulse of radiation converging on the spacetime point $x= 0$, 
and disappearing as  a flux-carrying particle into the extra dimension.
This situation would be classically improbable. 

We will therefore set in (\ref{strengthphi}) 
\begin{equation}
\varphi= g \, G_{R}.
\end{equation}

\section{First integral of the equations of motion} \label{sec3}

The equations of motion for the electric charge in the background field of the magnetic event are given by the standard Lorentz force,
\begin{equation} \label{Lorentzforce}
 m \frac{d^2 x^{\mu}}{ds^2}= q F^{\mu \nu} \frac{d x_{\nu}}{ds}.
\end{equation}
where $  F^{\mu \nu}$ is the field given in the previous section.

Because the field of the event is invariant under Lorentz transformations the three components of the spacetime angular momentum,
\begin{equation} \label{J2}
J^{\rho}=\frac{1}{2} \epsilon ^{\rho \mu \nu} J_{\mu \nu},
\end{equation}
are conserved. Just as it happens for an ordinary magnetic pole in three space dimensions (see for example \cite{Coleman}, and Appendix D of \cite{BunsterP}),   the angular momentum Eq. (\ref{J2})  has a ``spin"  piece additional to the orbital part. As discussed in the Appendix, when the electric charge is within the future light cone of the magnetic event,  we have 
\begin{equation} \label{total}
J_{\mu}= L_{\mu}+S_{\mu},
\end{equation} 
with 
\begin{equation}
L_{\mu}=\epsilon_{\mu \nu \rho} x^{\nu} m \frac{d x^{\rho}}{ds}
\end{equation}
and
\begin{equation}
S_{\mu}=- \kappa \frac{x_{\mu}}{\sqrt{-x^\alpha x_\alpha}}.
\end{equation}
Here 
\begin{equation}
\kappa= \frac{ q g}{ 2 \pi}
\end{equation}
and $s$ is the proper time. On the other hand, when the electric charge is not under the influence of the magnetic event one has just $J_{\mu}=L_{\mu}$.

We can always choose a Lorentz frame in which, before reaching the future light cone of the event,  the electric charge is at rest, at 
 $x^1=r_0, x^2=0$. Then we have 
\begin{eqnarray} \label{initial}
J_0&=&0, \\ 
J_1&=&0,\\
J_2&=&-m r_0,
\end{eqnarray}
where we have set $\epsilon_{012}=1$.  

The value of $J_{\mu}$ given by  Eqs. (12)--(14), which was arrived at by calculating it before the interaction begins, must also be valid afterward since the total angular momentum is conserved. As we shall see in Sec. V, during the scattering process there will be an exchange of spacetime orbital angular momentum and spin so that the sum $J_{\mu}$ remains constant.

It follows from Eqs.(12)--(14) that the conserved total angular momentum $J_{\mu}$ is a spacelike vector. This is a consequence of the fact that the worldline of the electrically charged particle is timelike. One could only arrange for a null, or timelike,  $J_{\mu}$ by having a spacelike worldline for an electric charge, but this would not correspond to a physical particle.

Now, after the particle has entered the future light cone of the event, one has
\begin{equation} \label{uno}
J_{\mu} x^{\mu}= \frac{-\kappa x^{\alpha} x_{\alpha}}{\sqrt{-x^{\alpha} x_{\alpha}}}=\kappa \sqrt{-x^{\alpha} x_{\alpha}}
=\kappa \sqrt{t^2-x^2-y^2},
\end{equation}
where we have denoted $(x^0, x^1, x^2)= (t, x, y)$.

Since $J_{\mu}$ is conserved and only $J_2 \neq 0$, we have
\begin{equation}
J_2 y= \kappa \sqrt{t^2-x^2-y^2},
\end{equation}
or
\begin{equation}  \label{y}
y= a \sqrt{t^2-x^2-y^2}
\end{equation}
with
\begin{equation}
a= -\frac{\kappa}{m r_0}.
\end{equation}

\begin{figure} [b]
\centering 
\includegraphics[angle=0,width=6.0cm]{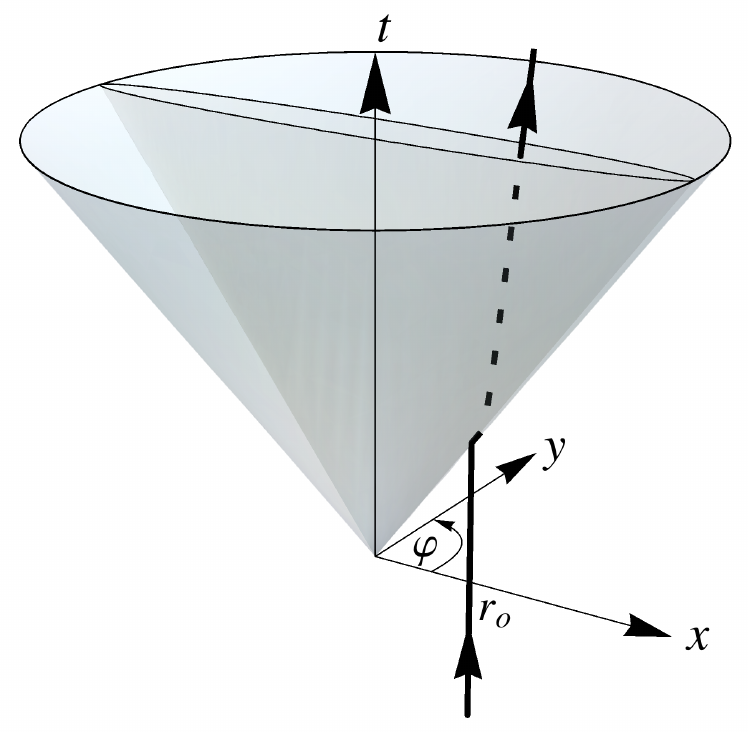} 
\caption{Worldline of the charged particle in the field of a magnetic event. The particle comes freely from the past and experiences a jolt when it crosses the future light cone of the magnetic event.  After this crossing, the particle spirals up in time, lying on the surface of a conical ellipsoid inscribed in the light cone of the event. Finally, the particle approaches an asymptotically straight worldline defined by $\varphi=\varphi_{\infty}$. In this graph $a=0.1$. } 
\label{3d} 
\end{figure} 

Equation (\ref{y}) may be written in parametric form as
\begin{eqnarray}
x&=& t \cos \varphi, \label{px}\\
y&=& \frac{a t}{\sqrt{1+a^2}} \sin \varphi, \label{py}
\end{eqnarray}
where $\varphi$ is the polar angle in the $x-y$ plane.

Geometrically these equations describe half of the ellipsoid
\begin{equation}
y^2(1+a^2)+a^2 x^2= a^2 t^2,
\end{equation}
which is inscribed in the light cone of the event, as illustrated in Fig. 1. The half $y\geq 0$ corresponds to $a>0$, and the half $y\leq 0$ to $ a<0$.

Next we proceed to evaluate $L_{\mu}$ and $S_{\mu}$ after the electric charge starts feeling the field of the event. First we observe from (\ref{y}) and (\ref{py}) that 
\begin{equation}
\sqrt{-x^{\mu} x_{\mu}}= \frac{y}{a}=\frac{t \sin \varphi}{\sqrt{1+a^2}}.
\end{equation}
Then
\begin{equation}
S_{\mu}= \frac{-\kappa x_{\mu} \sqrt{1+a^2}}{t \sin \varphi}.
\end{equation}
On the other hand, Eqs. (\ref{px}) and (\ref{py}) yield for the orbital angular momentum 
\begin{eqnarray}
L_0&=& \frac{ma}{\sqrt{1+a^2}} t^2 \frac{d \varphi}{ ds}, \\
L_1&=& -\frac{ma}{\sqrt{1+a^2}} t^2 \cos \varphi \frac{d \varphi}{ ds}, \\
L_2&=&- m t^2 \sin \varphi \frac{d \varphi}{ ds}.
\end{eqnarray}
Now we demand that the total angular momentum be conserved. Recalling Eqs. (\ref{initial})--(14), the conservation equations read 
 \begin{eqnarray}
\!\!\!\! J_0&\!\!=\!\!& -\kappa \left( -\frac{\sqrt{1+a^2}}{\sin \varphi}+\frac{1}{r_0 \sqrt{1+a^2}} t^2 \frac{d \varphi}{ ds} \right)=0,
\label{j0} \\
\!\!\!\! J_1&\!\!=\!\!&-\kappa \left( \frac{\sqrt{1+a^2} }{\sin \varphi} -\frac{1}{r_0\sqrt{1+a^2}} t^2  \frac{d \varphi}{ ds} \right) \cos \varphi =0, \label{j1}\\
\!\!\!\! J_2&\!\!=\!\!&-\kappa \left( a- \frac{ t^2 \sin \varphi}{a r_0} \frac{d \varphi}{ ds}\right)= -mr_0= \frac{\kappa}{a}. \label{j2}
\end{eqnarray} 
Equation (\ref{j2}) may be rewritten as 
\begin{equation} \label{equiv}
t^2 \sin \varphi \frac{d \varphi}{ ds}= r_0 (1+a^2).
\end{equation}
If  (\ref{equiv}) is inserted in (\ref{j0}) and (\ref{j1}), the latter two equations are also satisfied. Therefore  relation (\ref{equiv}) captures the full content of the angular momentum conservation.

Now
\begin{eqnarray}
ds^2&=&dt^2-dx^2-dy^2 \nonumber \\  &=& dt^2-(dt \cos \varphi-t \sin \varphi  d\varphi)^2 \nonumber \\&-&\frac{a^2}{1+a^2}(dt \sin \varphi+t \cos \varphi  d\varphi)^2,
\end{eqnarray}
which, combined with (\ref{equiv}), yields 
\begin{eqnarray} \label{dift}
&&\left[\left(\frac{t^2}{r_0^2}\frac{1}{1+a^2} +1\right) \sin^ 2 \varphi +a^2\right]  \left( t\frac{d \varphi }{dt}\right)^2 \nonumber \\ &&=  \sin^2 \varphi +2 t \frac{d \varphi }{dt} \cos \varphi \sin \varphi . 
\end{eqnarray}
Introducing the dimensionless parameter $\xi$ through 
\begin{equation}
t= r_0 e^{\xi},
\end{equation}
which implies $\displaystyle t \frac{d \varphi}{dt}= \frac{d \varphi}{d \xi}$, we can rewrite Eq. (\ref{dift}) as
\begin{equation} \label{segundo}
\frac{d \varphi }{d\xi}=\frac{\sin \varphi}{A^2}\left( \cos\varphi + \sqrt{\cos^2 \varphi+A^2}\right),
\end{equation}
with
\begin{equation}
A^2(\varphi,\xi;a)=\left(\frac{e^{2\xi}}{1+a^2} +1\right) \sin^ 2 \varphi +a^2.
\end{equation}
The choice of a plus sign in (\ref{segundo}) comes from (\ref{equiv}) which show us that $d \varphi$ and $\sin \varphi$ have the same sign.

Equations (\ref{px}), (\ref{py}) and (\ref{segundo}) constitute a first integral of the equations of motion.

\section{Complete integration of the equations of motion} \label{sec4}

We will exhibit in this section numerical results for the solution of the first-order differential equation (\ref{segundo}). The physical  situation is that the particle is free up to the time $t=r_0$,  when it is first hit by the field of the event, and afterward it continues to move under its influence. Now, if one were to assume--as one normally does--that the position of the particle is continuous, one would have that the solution of Eq. (\ref{segundo}) would be  $\varphi=0$ for all times. However, if one introduces that result in Eqs. (\ref{px}) and (\ref{py}), it follows that $x=t$, $y=0$ , i.e., the particle suddenly starts moving with the speed of light. This unphysical result stems from the fact that the Green function (\ref{Green}), and the field strength derived from it, diverge as one approaches the light cone from inside. To extract physically sensible results one must smear the source so that the field is regularized on the light cone, as illustrated in Fig. 2, and then extract conclusions which are insensitive to the details of the regularization. 

\begin{figure} [t]
\centering 
\includegraphics[width=0.45\textwidth]{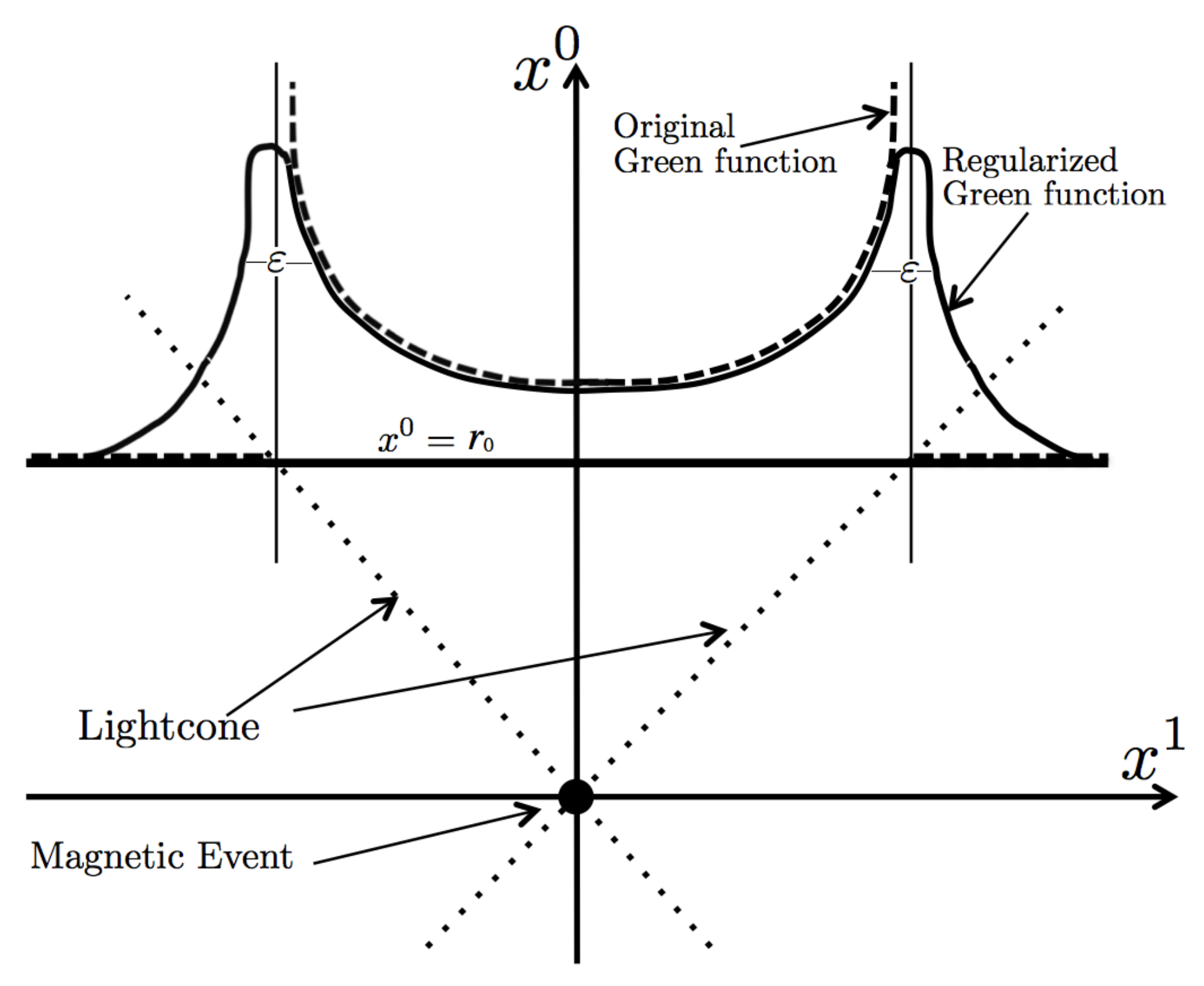} 
\caption{Regularization of the retarded field. A section of spacetime, the $(x^0,x^1)$ plane at $x^2=0$, is shown.  The magnetic event is at the origin. The original retarded Green function  $G_R$ is nonvanishing inside its future light cone only, but  diverges on it. After regularization, which may be thought of as the smearing of the magnetic event, the field becomes finite on the light cone, but has a rapidly decaying tail outside of it. One recovers the original Green function when the cutoff width $\varepsilon$ approaches $0^{+}$. } 
\label{greenregular} 
\end{figure}

The need for the regularization stems from the step-functions in the Green 
function (\ref{Green}). It is distinctly brought into evidence if one attempts to apply the Gauss law for a spacetime region which contains the magnetic event,  and whose boundary is formed by two spacelike disks and a timelike cylinder that joins them. The first spacelike disk is taken to be on  the future of the event and it extends slightly outside of its light cone, similarly for the second disk, but to the past of the event. If one takes for ${}^{*}F_{\mu}$ in (\ref{strengthphi}) the one derived from the Green function (\ref{Green}), one finds that its flux across the upper disk diverges whereas the contribution of the
lower disk and of the cylinder vanish. Thus, the total flux over the closed surface enclosing the event diverges. However, on account of Eq. (\ref{field})  the Gauss formula yields that the flux should be equal to the magnetic charge $g$. The regularization resolves the problem by smoothing out the Green function, so that the Gauss formula can be properly applied. In the limit $\varepsilon \to 0^{+}$ for the regulator,  the flux across the upper disk becomes $g$, rather than infinity, while the contribution of  the cylinder vanishes (the contribution of the lower disk vanishes already before the limit).

Thus we will assume that the regularization has been implemented, that the particle has crossed the light cone, and will start the integration assuming that $\varphi (t=r_0)=\varphi (\xi =0)=\varphi(0)$ is different from zero, but very small. This is  because,  due to the regularization, the particle starts experiencing a force a short time before hitting the light cone, and therefore when $t=r_0$ it has already turned a little bit. 

\begin{figure}[b] 
\centering 
\includegraphics[angle=0,width=0.40\textwidth]{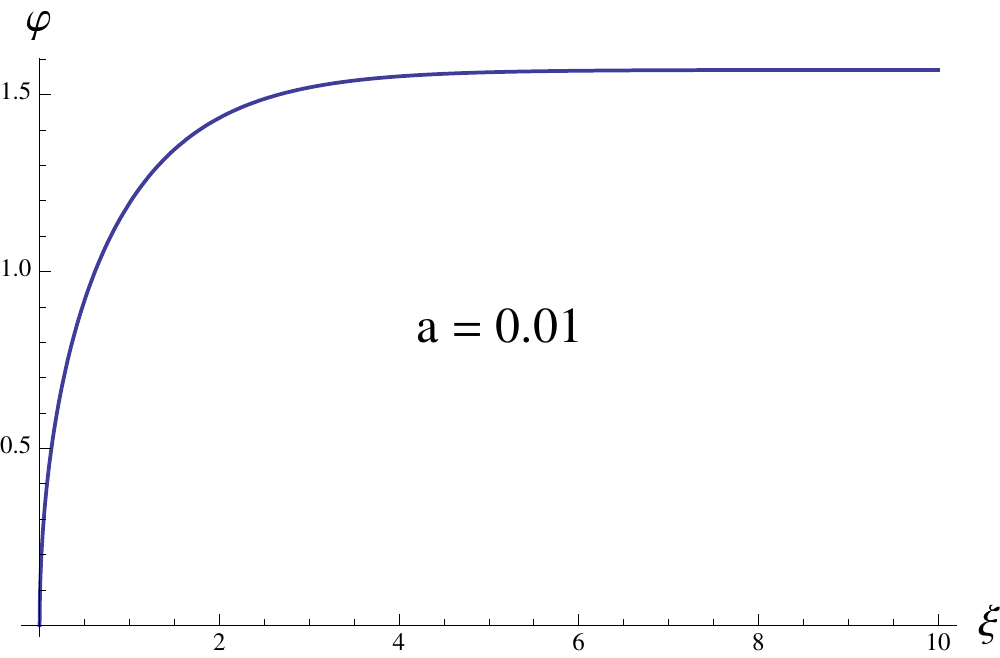} 
\vskip 0.5cm
\includegraphics[angle=0,width=0.40\textwidth]{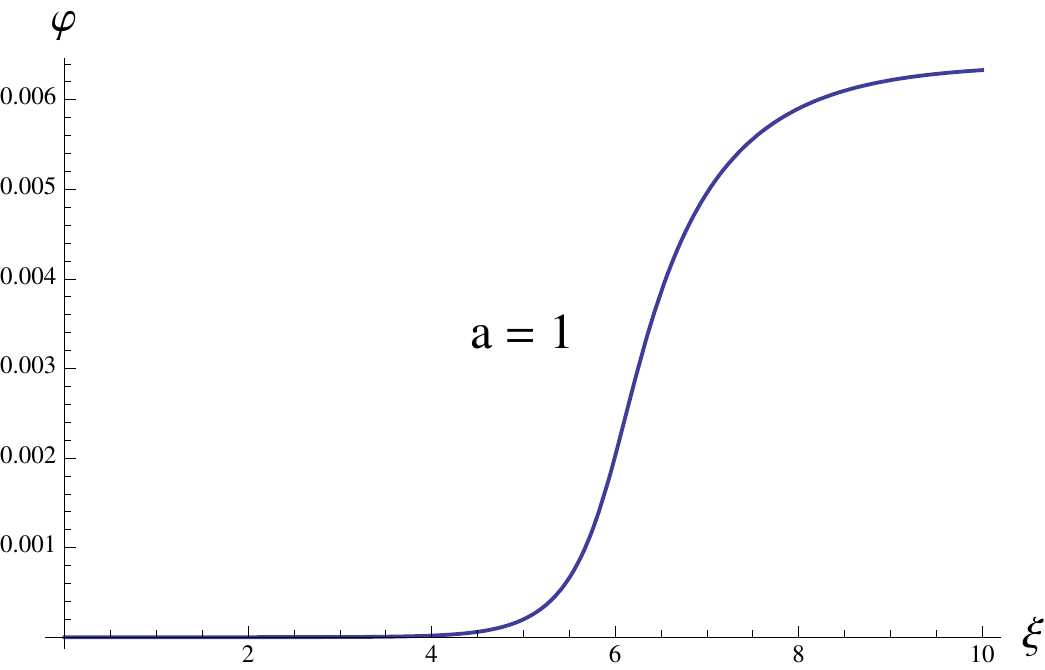} 
\caption{Time dependence of the polar angle. The graphs show   $\varphi$ as a function of  $\xi=\ln(t/r_0)$ for the cases  $a=0.01$ and $a=1$. In both cases  $\varphi(\xi)$ approaches asymptotically  a constant value $\varphi_{\infty} < \pi/2$. The cutoff was implemented by taking $\varphi(0)=10^{-9}$.  } 
\label{varphi} 
\end{figure} 

Note that the regularization procedure changes the electromagnetic field appearing on the 
right-hand side of the equation of motion (\ref{Lorentzforce}) to 
\begin{equation}
F_{\mu \nu}= \epsilon_{\mu \nu \rho} \partial^{\rho} g G_{\textrm{R}}^{\textrm{regularized}}.
\end{equation}
Therefore due to the antisymmetry of $F_{\mu \nu}$, one still has, after regularization,  
\begin{equation}
\frac{d}{ds}( v_{\mu}v^{\mu})=0.
\end{equation}
Hence, if the worldline of the particle was timelike before reaching the (smoothed) future light cone of the event, it continues being so while it crosses it and also afterward, inside of it.

\begin{figure} [b]
\centering 
\includegraphics[angle=0,width=0.4\textwidth]{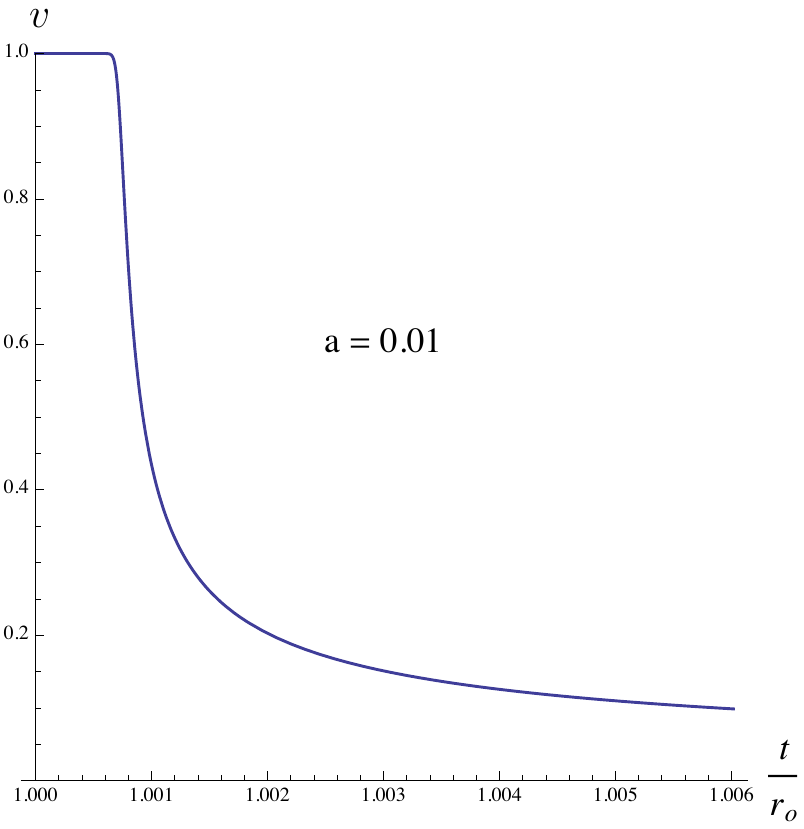} 
\vskip 0.5cm
\includegraphics[angle=0,width=0.4\textwidth]{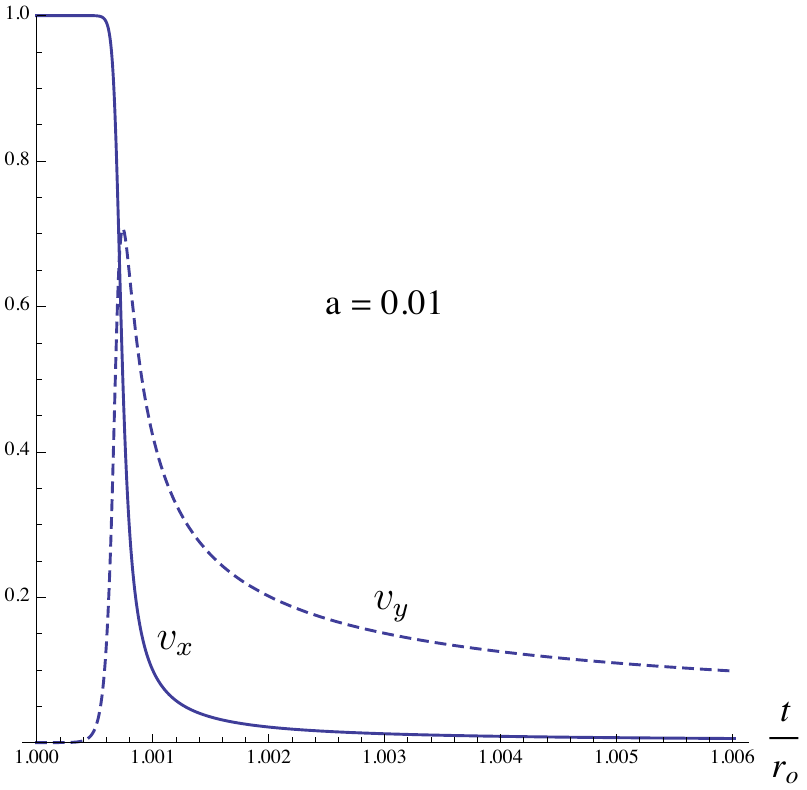} 
\caption{Particle velocity as a function of time. The first graph shows the velocity $v$ of the particle as a function of  $t/r_0$ for the case  $a=0.01$. The second graph shows the time dependence of the components $v_x$ and $v_y$ and exhibits the sudden turn and coming to rest of the particle. The cutoff was implemented by taking $\varphi(0)=10^{-9}$. } 
\label{velog} 
\end{figure}

\subsection{ Limit $\xi \rightarrow \infty$ }

There is one conclusion that one can obtain right away, before a numerical analysis. It is the asymptotic behavior for large times.

The asymptotic form of Eq. (\ref{segundo}) is
\begin{equation}
\frac{d\varphi}{d \xi}=\sqrt{1+a^2} e^{-\xi}+O(e^{-2\xi}), 
\end{equation}
yielding
\begin{equation} \label{asy}
\varphi( \xi) \sim \varphi_{\infty}-\sqrt{1+a^2} e^{-\xi}, 
\end{equation}
when  $\xi \rightarrow \infty$. Here $\varphi_{\infty}$ is a constant. This implies that for $t\rightarrow \infty$, the trajectory of the particle is a straight line, with
\begin{equation}
y= \frac{a x \tan \varphi_{\infty}}{\sqrt{1+a^2}}.
\end{equation}

Moreover, taking in account  Eq. (\ref{asy}) it is possible to find the velocity of the particle for large times. The asymptotic expression for the velocity is given by
\begin{equation} \label{velocityasy}
v_{\infty}=\sqrt{\frac{\cos^2 \varphi_{\infty}+a^2}{1+a^2}}.
\end{equation}

\subsection{Numerical integration}

We now exhibit through a sequence of figures the dependence of the solution of the equation of motion on the spacetime impact parameter  $r_0$, or rather its dimensionless 
inverse $a= -q g/ (2 \pi m r_0)= -\kappa /(m r_0)$. Special attention is paid to the spacetime scattering angle $\varphi_{\infty}$.  The qualitative behavior is independent of the regularization length $\varepsilon$. After crossing the light cone the particle starts spiraling up in time and it settles for large times into uniform motion according to the discussion of the previous subsection. The angle $\varphi_{\infty}$ is an increasing function of the  impact parameter. The precise form of the function is cutoff dependent but its asymptotic value for $r_0\rightarrow \infty$ is cutoff independent and equal to $\pi/2$. Figures 3 and 4 are drawn for a fixed value of the cutoff  $\varphi(0)=10 ^{-9}$. Figure 3 exhibits the time dependence of the polar angle $\varphi$ for a small value of $a$ and for a large one. Figure 4 exhibits the time dependence of the velocity and its components for $a=0.01$.  Figure 5 illustrates  the dependence of the scattering angle  $\varphi_{\infty}$ on the inverse impact parameter $a$, for different values of the cutoff,  and it shows that, for small $a$,  $\varphi_{\infty}= \pi/2$, independently of the cutoff.

\section{Conclusion} \label{conclusion}

The motion in the frame in which the electric charge is initially at rest is as follows: As the particle crosses the light cone it starts rotating and slowing down. The rotation is counterclockwise for $a>0$ and clockwise for $a<0$ [recall Eq. (\ref{py})]. This is in agreement with the fact that under space reflection $a$ is a pseudoscalar because  the product $q g$ of the electric and magnetic charges is a pseudoscalar. 

The limit $a\to 0$ corresponds to a large impact parameter and it is of particular interest because one would expect that then the details of the magnetic source, idealized here to be a point, should be irrelevant. As seen in Fig 1, when $a \rightarrow 0$, the inscribed ellipsoid becomes very narrow and parallel to the $x$ axis. Hence, after a time of the order of the impact parameter,  the particle makes sharply a quarter of a turn and comes to rest at the same spatial position at which the event happened in the past. Thus the net effect of the scattering is to change the state of the particle from being at rest $x=r_0$, $y=0$ at $t=r_0$, to being again at rest  $x=0$ , $y=0$ for $t \gg r_0 \gg qg/(2 \pi m)$.  

This jump is the main signature of the presence of the magnetic event as felt by an electric charge.

As was anticipated in Sec. III, we note that during the scattering process there is an exchange of spacetime orbital angular momentum and spin. Indeed in the final stage when the particle is at rest above the event, after a very long time, it has a nonvanishing $S_0= \kappa$ which is compensated by $L_0=-\kappa$. On the other hand, initially one has $L_0=S_0=0$.

\begin{figure} [t]
\centering 
\includegraphics[angle=0,width=0.45\textwidth]{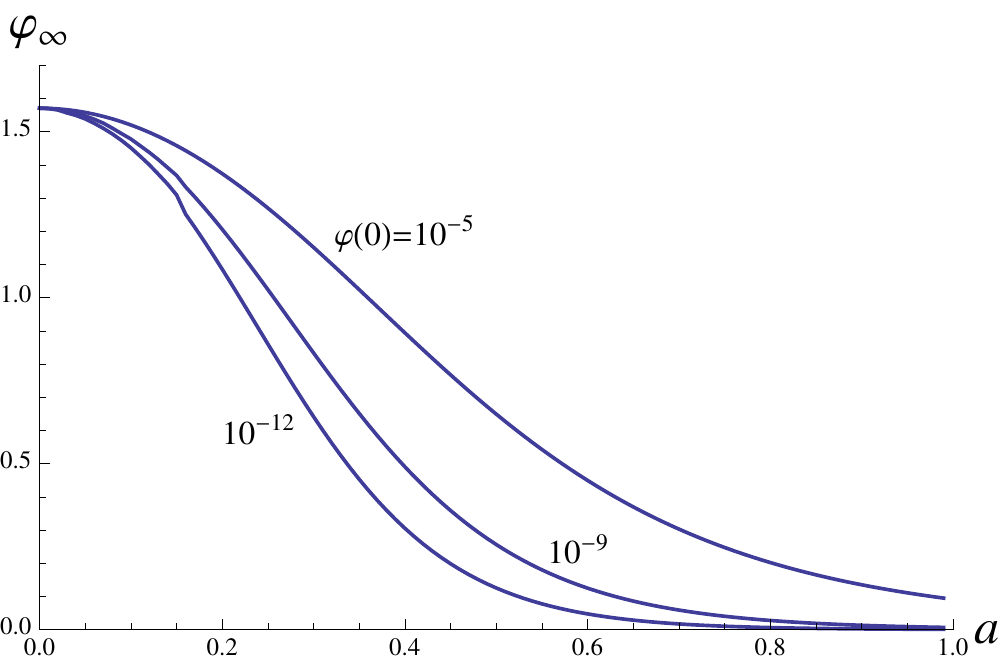} 
\caption{Asymptotic  value  of the spacetime scattering angle $\varphi_{\infty}$ as a function of the inverse of the impact parameter, $a$. In the case of a strong interaction ($a$ large), the graph shows that $\varphi_\infty$ tends to zero. For  small $a$, i.e., the case of a weak interaction,  $\varphi_\infty$ approaches $\pi/2$.} 
\label{Bfig} 
\end{figure} 

\acknowledgments
Appreciation is extended to Constantin Bachas and Marc Henneaux for illuminating discussions on the subject of charged events, and to A. Gomberoff for Fig. 2. This work was partially funded by Fondecyt Grants No. 1085322, 1095098, 1100755, and by the Conicyt Grant No. ACT-91. The Centro de Estudios Cient\'{\i}ficos (CECs) is funded by the Chilean government through the Centers of Excellence Base Financing Program of Conicyt.  

\appendix

\section{Angular momentum in the field of a magnetic event}

We work in the magnetic representation, in which a Dirac string is attached to the electric charge. The worldsheet $\Sigma$ of the string will be denoted by $x^{\mu}=y^{\mu}(\sigma,\tau)$, with $y^{\mu}(0,\tau)=z^{\mu}(\tau)$, the worldline of the electric charge. 

The action for the electric charge in a background field $F_{\mu \nu}$ is given by
\begin{equation} \label{act}
I= I_{\textrm{kinetic}}+I_{\textrm{interaction}},
\end{equation}
with
\begin{equation} \label{kinetic}
I_{\textrm{kinetic}}= -m \int_{-\infty}^{+\infty}d \tau \sqrt{-\dot{z}^{\alpha}\dot{z}_{\alpha}},
\end{equation}
and
\begin{eqnarray}
I_{\textrm{interaction}}&=& \frac{q}{2} \int_{\Sigma}  F_{\mu \nu}(y) dy^{\mu} \wedge dy^{\nu} \nonumber \\
&=&q \int_{\Sigma}  F_{\mu \nu}(y) \frac{\partial y^{\mu}}{\partial \sigma} \frac{ \partial y^{\nu}}{\partial \tau} d\sigma d\tau.
\end{eqnarray}

If one varies $y^{\mu}$ assuming that the string does not cross a magnetic source (``Dirac veto"), so that $d F=0$, one finds by using the Stokes theorem that
\begin{equation}
\delta I_{\textrm{interaction}}= -q \int_{-\infty}^{+\infty}  F_{\mu \nu}(z) \frac{d z^{\mu}}{d \tau} \delta  z^{\nu}( \tau) d\tau,
\end{equation}
which yields the Lorentz force, as desired.

If one takes $ F_{\mu \nu}(x)$ to be the field of an event at the origin, one has, from Sec. \ref{sec2},
\begin{equation} \label{Iinterac}
I_{\textrm{interaction}}= \frac{q g}{4 \pi} \int \epsilon_{\mu \nu \lambda} \frac{y^{\lambda}}{(-y^2)^{3/2}}   dy^{\mu} \wedge dy^{\nu},
\end{equation}
where $y^2 = y^{\alpha} y_{\alpha}$. The integral is extended over the intersection of the worldsheet $\Sigma$ with the interior of the future light cone of the event, since the field vanishes outside that cone.

Now, $I_{\textrm{interaction}}$ is manifestly Lorentz invariant and so is  $I_{\textrm{kinetic}}$. Therefore we may follow the standard Noether procedure for the transformation,
\begin{equation} \label{LT}
\delta y^{\nu}= \epsilon^{\nu \alpha \rho} y_{\alpha} \xi_{\rho},
\end{equation}
which is an infinitesimal Lorentz rotation with parameter $ \epsilon^{\nu \alpha \rho}  \xi_{\rho}$.

The conserved total charge is then
\begin{equation}
J_{\rho}=L_{\rho}+S_{\rho},
\end{equation}
where $L_{\rho}$ is the standard orbital angular momentum coming from $I_{\textrm{kinetic}}$ and $S_{\rho}$ is given by
\begin{equation}
 \xi_{\rho} S^{\rho}= \int_{0}^{\infty} \frac{\delta {L_{\textrm{interaction}}}}{\delta \dot{y}^{\mu}(\sigma)}\delta y^{\mu}(\sigma) d\sigma,
\end{equation}
where the interaction Lagrangian $L$ is given by
\begin{equation}
 L_{\textrm{interaction}}=\frac{q g}{2 \pi} \int_{0}^{\infty} \epsilon_{\mu \nu \lambda} \frac{y^{\lambda}}{(-y^{\alpha} y_{\alpha})^{3/2}}   \frac{\partial y^{\mu}}{\partial \sigma} \frac{ \partial y^{\nu}}{\partial \tau} d\sigma,
\end{equation}
so that its integral over $\tau$ gives the action (\ref{Iinterac}). 

One obtains
\begin{equation} \label{ss}
S^{\rho}= \frac{qg}{2 \pi} \int_{0} ^{\infty} d\sigma \left( \frac{1}{2} \frac{\partial y^2}{\partial \sigma} y^{\rho}-\frac{\partial y^{\rho}}{\partial \sigma} y^2 \right) ( -y^2) ^{-3/2}.
\end{equation}
But the integrand in (\ref{ss}) is just $ d(y^{\rho} (-y^2)^{-1/2})/ d \sigma$. Therefore (\ref{ss}) yields,
\begin{equation}
S^{\rho}=\frac{q g}{2 \pi} \left[ \frac{y^{\rho}}{(-y^2)^{1/2}} \left|_{\sigma=\infty} \right. - \frac{z^{\rho}}{(-z^2)^{1/2}}\right].
\end{equation}

The contribution at spatial infinity vanishes after one regularizes the field as illustrated in Fig. 2. So one finds the expression given for $S_{\rho}$ in the main text, Eq. (10).

One may view  the spin $S^{\rho}$ as stemming from a ``rotational inertia" acquired by the self-field of the electric charge moving through the background field of the event. Direct calculation of the angular momentum stored in the field \cite{BBH} sustains this interpretation.

\end{document}